\documentclass[10pt,twocolumn]{article} 
\usepackage{simpleConference}
\usepackage{times}
\usepackage{graphicx}
\usepackage{amssymb}
\usepackage{url,hyperref}

\begin{document}

\title{Analysis and Modeling of Behavioral Changes in a News Service}

\author{Atom Sonoda \\
\\
The University of Tokyo \\
Bunkyo-ku, Tokyo, Japan \\
\today
\\
\\
sonoda@crimson.q.t.u-tokyo.ac.jp  \\
}

\author{Fujio Toriumi \\
\\
The University of Tokyo \\
Bunkyo-ku, Tokyo, Japan \\
\today
\\
\\
tori@sys.t.u-tokyo.ac.jp  \\
}

\author{
Atom Sonoda, Fujio Toriumi \\
The University of Tokyo \\
Bunkyo-ku, Tokyo, Japan \\
\today\\
sonoda@crimson.q.t.u-tokyo.ac.jp, tori@sys.t.u-tokyo.ac.jp  \\
\\
Hiroto Nakajima, Miyabi Gouji \\
Nikkei Inc. \\
Chiyoda-ku, Tokyo, Japan \\
\{hiroto.nakajima, miyabi.goji\}@nex.nikkei.com  \\
}

\maketitle
\thispagestyle{empty}

\begin{abstract}
Information is transmitted through websites, and immediate reactions to various kinds of information are required. Hence, efforts by users to select information themselves have increased, which is fueling further improvements in recommendation services that can reduce such burdens. On the other hand, filter bubbles that only provide biased information to users are generated due to redundant recommendations. In this research, we analyzed behavioral changes prior to recommendation by clustering, and we found that user attributes and cluster contents are different among users with different behavioral changes. The proportion of users under forty and women was relatively large in the diversity-increasing group.

We also proposed an article selection model to clarify the influence of recommendation systems on behavioral changes. We compared our proposed model with the target data, verified it, and evaluated the effect of recommendation systems on user behavior. Our simulation results showed that diversity usually decreases, but collaborative filtering can suppress the diversity decrease more effectively than non-recommendations. We also found that the category that users are interested in the most is easily strengthened and is one factor that leads to less diversity, and a recommendation method that can suppress the strengthening of the category that users are interested in the most will be effective for developing a recommendation system that can suppress diversity decreasing.
\end{abstract}

\section{Introduction}
\hspace{1em}
Information is transmitted through websites and newspapers, and immediate reactions to various kinds of information are required. Hence, we believe that the efforts by users to select information themselves are increasing, which is leading to refinement of recommendation services that can reduce such burdens in many fields, especially for e-commerce and news services. Many companies, including Amazon, Netflix, and Apple, have also improved their recommendation systems to increase their profits \cite{schafer1999recommender}.

On the other hand, filter bubbles, which only provide biased information to users, are generated due to excessive recommendations \cite{pariser2011filter}.
People actively read specific information sources using the internet, and this practice may decrease selection diversity.
Recommendation systems for news articles might also change user selections \cite{mooney2000content} \cite{nguyen2014exploring}. 
A recommendation system must be developed that can provide opportunities to come in contact with various opinions and prevent filter bubbles.

In this research, we analyzed behavioral changes prior to developing recommendations by clustering and showed that behavior changes during a certain period.

In Section \ref{sec:clustering}, to analyze user behavior, we propose a method based on a user's browsing history to classify articles from online news services provided by Nikkei Inc., one of the largest newspaper companies in Japan. In Section \ref{sec:analysis_behav}, we analyze the process of behavioral changes using information on categorized articles. In Section \ref {sec:analysis_feature}, we show the characteristics of users whose browsing behavior changed.

In Section \ref{sec:modeling}, we propose an article selection model to clarify the influence of recommendation systems on behavior changes. Then we compare the results from a simulation and analysis from Sections \ref {sec:analysis_behav} and \ref{sec:analysis_feature} and verify the model.

\section{Related works}
\hspace{1em}
Along with the development of online media, research on recommendation systems is increasing.
Schafer et al.　\cite{schafer1999recommender}, Sarwar et al. \cite{sarwar2001item}, and Linden et al. \cite{linden2003amazon} analyzed the use of them in e-commerce from the viewpoint of marketing viewpoint and scalability. 

Schein et al. \cite{schein2002methods} considered a solution to the cold start problem, which is the difficulty of recommending items that were not evaluated sufficiently when introducing recommendation systems. Senecal et al. \cite{senecal2004influence} showed that online recommendation systems are more influential than recommendations from traditional human experts or other consumers.
In addition, Lee et al. \cite{lee2002intelligent} argued that appropriate recommendation systems depend on target products.

Pariser \cite{pariser2011filter} identified a filter bubble that only provides user-biased information by excessive recommendations.
Bakshy et al. \cite{bakshy2015exposure} analyzed news articles on Facebook and argued that cross-cutting news from information sources from the opposite political spectrum is not shared. Because of recommendation systems, articles are selectively presented in the field of political news encountered in social media.

On the other hand, contrary to Pariser's conclusion, Linden \cite{pariserWrong} argued that since people have difficulty obtaining exposure to ideas and articles of which they are unaware, recommendation systems increase serendipity.

Fleder et al. \cite{fleder2009blockbuster} discussed the impact of recommendation systems on sales diversity and argued that some recommendations lead to a net decrease in such diversity, but personal diversity increases. However, this model is limited to two items, and the process of adding new items like news articles was ignored.

Nguyen et al. \cite{nguyen2014exploring} showed that diversity of both recommended and evaluated items decreases after a certain period of time in MovieLens\footnote{http://www.movielens.org/}, however the diversity of items, whose users who are affected by the recommendation system using collaborative filtering, decreases only a little.
However, analysis is insufficient of the biases of users who are subjected to filtering and those who are beyond its affects, and only collaborative filtering is handled as a recommendation system.

In addition, Cosley et al. \cite{cosley2003seeing} concluded that the interface of recommendation systems influences the acceptance of recommendations. Ahmed et al. \cite{ahmed2014agent} evaluated the performance of real-time searches on twitter by simulations.

Although these previous studies present interesting knowledge about filter bubbles, behavioral changes in online media, and the development of recommendation systems and their influences, discussion remains insufficient about media that are relatively impartial like most newspaper companies\cite{kiousis2001public}. Comparisons among recommendation systems are also inadequate. 
Therefore, in this research, we analyze behavioral changes in online news media and compare content-based recommendation systems and collaborative filtering based on user's similarity and show how recommendation systems work in online media.

\section{Network clustering method\label{sec:clustering}}

\subsection{Dataset}
\hspace{1em}
In this research, we use data from online news services provided by Nikkei Inc. during a three-month period from May 21 to August 20, 2017. About 600,000 articles were read during this time, and the number of subscribers was about 2 million.

We excluded articles that were read by fewer than 100 users for two reasons. 
First, in the development of future recommendation systems, a system should recommend articles that are as new as possible because such systems are designed for news services that demand immediacy and periodic updates to reduce the amount of data from the viewpoint of calculation time.
Second, to prevent filter bubbles in recommendations, a system should not recommend esoteric article that will only appeal to few users.
After conducting these processes, about 70,000 articles remained for analysis.

\subsection{Article network construction}
\hspace{1em}
In this section, we explain a method that classifies the articles to grasp the change of user's interest. 
We classified articles using the clustering method proposed by Baba et al. \cite{baba2015classification} \cite{uchida2017evaluation}. Although the method was proposed to classify twitter posts, we applied it to classify the articles. With it, we can extract the topics of articles based on a user's browsing history without language information. 
We summarize this network clustering method as follows. We classified articles from Nikkei news based on the similarity of the users who read them by calculating the similarity between a pair of articles. This similarity is based on overlapping users who read both articles, and we constructed an article network using the similarity. 
By classifying the articles, we can evaluate s user's behavioral changes by clusters.

When two or more users read the same two articles, the two articles have common interests. In other words, the similarity of the articles can be calculated from the degree of overlap of the users who read them. Therefore, we can construct an article network by linking articles with high similarity. The degree of the similarity of two articles, $a_{i}$, $a_{j}$, is calculated using Simpson coefficients:
\begin{eqnarray}
Sim(a_{i}, a_{j})=\frac{|U_{i} \cdot U_{j}|}{min(|U_{i}|,|U_{j}|)}.
 \label{eq:sim_kiji}
\end{eqnarray}
In this formula, $U_ {i}$ and $U_ {j}$ represent the user groups of articles $ a_ {i} $ and $ a_ {j}$.
We linked articles with a similarity of 0.62 or more and constructed a weighted network, which has modularity Q = 0.936, 38002 links, and 17,233 nodes.

Other indices measure similarity as well as the Simpson coefficient, e.g., the Jaccard coefficient and the Dice coefficient. Baba et al. used the Jaccard coefficient for similarity, but we applied the Simpson coefficient because it appropriately expresses the degree of relationships based on co-occurrence \cite{matsuo2005social}.

\subsection{Network clustering\label{subsec:clustering}}
\hspace{1em}
Next, we classify the above article network and acquire a set of similar articles. For community detection, we use the Louvain method \cite{blondel2008fast} based on modularity, which represents the degree of connectivity among a set of clusters. Highly connected clusters can be detected by maximizing modularity.

\begin{figure} [tbp]
\begin{center} 
\includegraphics[width=0.8\linewidth]{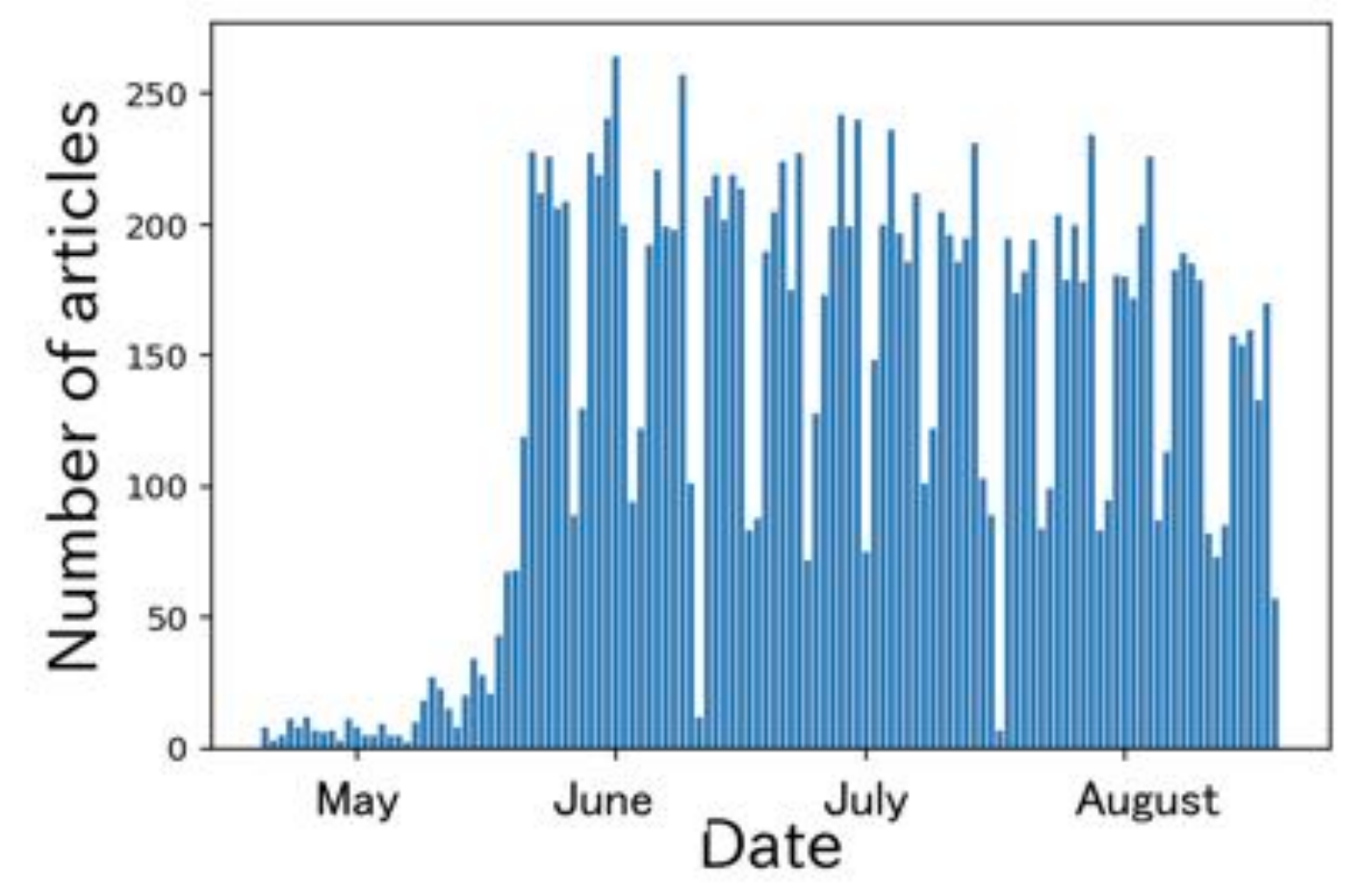} 
\caption{Trends in number of articles in one cluster}
\label{kiji_in_cluister}
\end{center} 
\end{figure}

The number of articles included in one of the clusters is presented in Fig. \ref{kiji_in_cluister}. The horizontal axis is the posted date of the articles, and the vertical axis is the number of articles. The amount of articles fluctuates depending on the day of the week, weekends, and holidays . 
This figure includes many articles published during the target period. On the other hand, the number of articles in the clustering result decreased in the later stage, since the number of days after publication is small and the amount of browsing is not properly accumulated. In this research, we compared the first and middle periods of the clustering target period.

\section{Analysis of behavioral changes\label{sec:analysis_behav}}

\subsection{Evaluation of diversity of behavior\label{subsec:eval_behav}}
\hspace{1em}
We defined the behaviors of the users as the class of articles consumed by them and evaluated their behavioral changes after a certain period of time. 
In this research, we analyzed the change of the diversity of an article's cluster by evaluating the user's behavior by the diversity of the cluster to which the article being read belongs.

We evaluated the browsing articles of each user based on the cluster to which the article belongs and the diversity of the browsing behavior based on the degree of the cluster concentration of read articles, calculated using information entropy:
\begin{eqnarray}
H(u)=-\sum_{i}p_{i} \cdot log p_{i}.
 \label{eq:entropy}
\end{eqnarray}
In this formula, $p_{i}$ represents the existence probability of each user of each cluster $c_{i}$.

A user with high cluster entropy is reading articles from various clusters, and a user with low cluster entropy is intensively concentrating on articles that just belong to a specific cluster.

We evaluated the diversity of browsing behaviors during the period and analyzed its changes. We also analyzed the characteristics of users whose diversity changed. 
Generally, with filter bubbles, recommendation systems isolate users from information that does not match their viewpoints, and the information is limited to a range of interest to the user.
We believe that influence of the recommendation systems can be measured by the cluster concentration degree of the articles being read. For example, if filter bubbles limit the contact of users to information that matches the interest espoused by the filter bubbles, the articles will be concentrated on a specific cluster, reducing diversity and decreasing cluster entropy.

\subsection{Experiment\label{experiment}}
\hspace{1em}
In our analysis, we compared the behaviors of June 1 to 10 (period 1) and July 6 to 15 (period 2) because the number of articles included in these clusters decreased in the later stage of the clustering target period (Fig. \ref{kiji_in_cluister}).

In this period, we selected users who joined on May and read 10 to 100 articles and extracted 1037 users who read at least one article that was included in the cluster. For these users, we calculated cluster entropy $H(u)$.

\subsection{Results and discussions}
\begin{figure} [ tbp]
\begin{center} 
\includegraphics[width=0.8\linewidth]{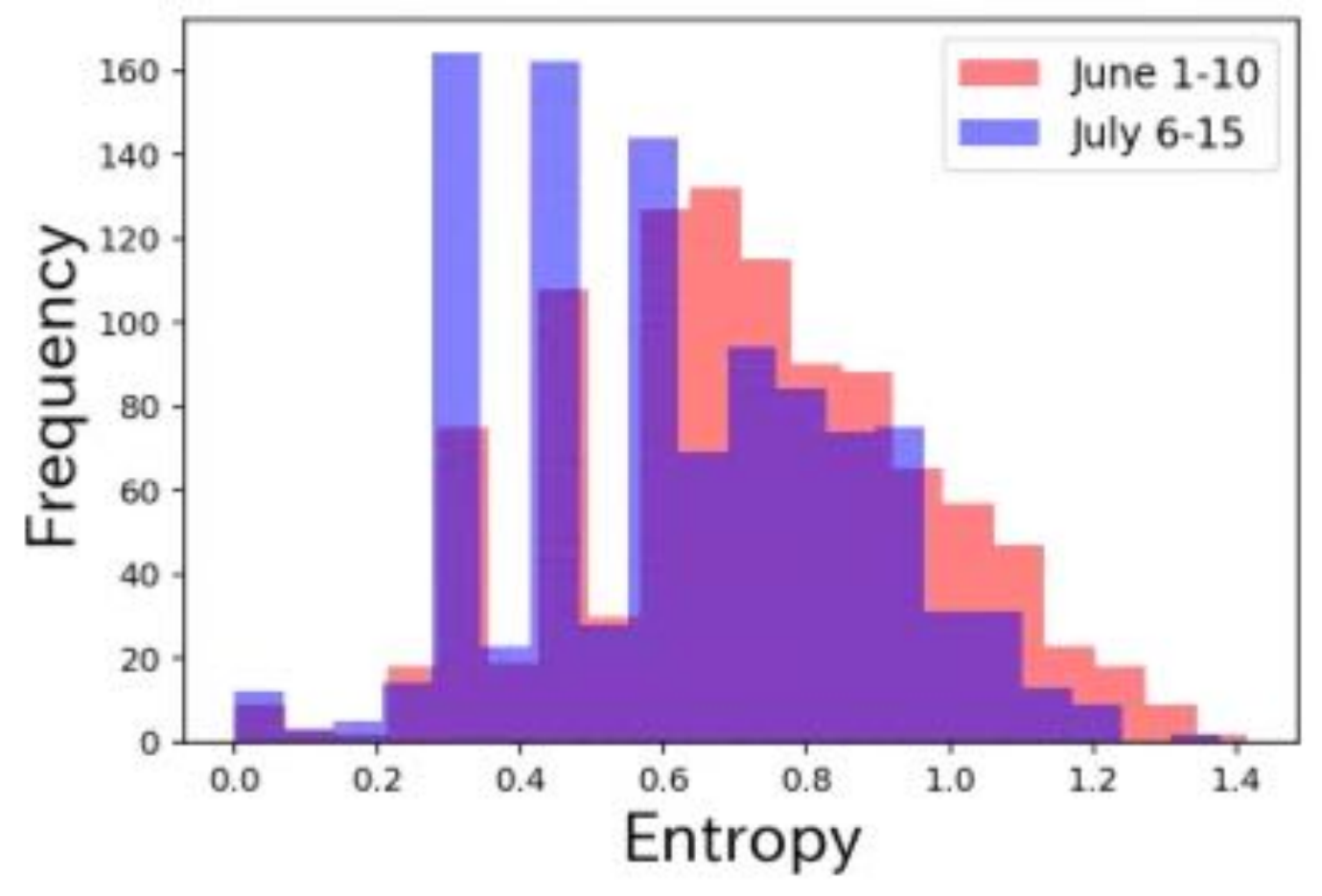} 
\caption{Change in cluster entropy}
\label{entropy}
\end{center} 
\end{figure}

\hspace{1em}
Figure \ref {entropy} shows the cluster entropy, which is an indicator of the degree of the concentration of clusters based on a user's interest and measures the range of the interest of articles read by users. 
In other words, a decrease in cluster entropy means that interest has narrowed.

The average cluster entropy in period 1 is 0.718, and it is 0.619 in period 2. The cluster entropy decreases significantly at a significance level of 1\%. 
Therefore, from this analysis, even in existing display systems, interest will be biased as the period continues. We believe the introduction of recommendation ranking systems is affecting this aspect.

\section{Analysis of features leading to behavioral change\label{sec:analysis_feature}}

\subsection{Features and target users}
\hspace{1em}
Next we analyzed the change of each user's cluster entropy of periods 1 and 2. As mentioned in the Section \ref{sec:analysis_behav}, cluster entropy decreased as a whole; that is, diversity decreased. On the other hand, since there were users whose cluster entropy and diversity increased, we compared a group whose diversity increased with a group whose diversity decreased.

The diversity-increasing group has 355 users out of 1037 users. For the diversity-decreasing group, we selected the same number of top 355 users as the increasing group from the diversity-decreasing group.
We compared such user attributes as age and gender as well as characteristic clusters of the diversity-increasing and diversity-decreasing groups.

\subsection{Attributes}
\hspace{1em}
The target data included information about prefecture, occupation, age, and gender as user attributes, each of which was divided into ten or more categories. Since insufficient users were included to confirm significant differences in each category, our analysis focused on age and gender.

Table \ref {feature} shows the results of each indicator. We found more women in the diversity-increasing group. When we divide users into 2 groups by ages are over 40 and ages are between 10 and 39, the number of users whose ages are between 10 and 39 of diversity-increasing group is bigger compared to the ones in diversity-decreasing group.

Looking at each generation, users in their 60s are more frequent in the diversity-decreasing group.

\begin{table} [ tbp]
\begin{center} 
\caption{Changes in averages of feature values
\newline
(NS means not significant)
}
\label{feature}
\begin{tabular} {|c|c|c|c|}
\hline
Diversity&Increasing&Decreasing&Significance level\\
\hline
gender (male/female)&(268, 87)&(298, 57)&P$<$0.01\\
\hline
(over 40, under 39)&(223, 132)&(268, 87)&P$<$0.05\\
\hline
age under 30&72&54&P$<$0.10\\
\hline
age 30&60&47&NS\\
\hline
age 40&87&83&NS\\
\hline
age in 50s&91&98&NS\\
\hline
age in 60s&32&54&P$<$0.05\\
\hline
age over 70&19&13&NS\\
\hline
\end{tabular} 
\end{center} 
\end{table}

\subsection{Characteristic clusters}
\hspace{1em}
Next we analyzed the clusters that are often included in one of the diversity-increasing or diversity-decreasing groups under the assumption that the behavior of users who consume articles in a specific cluster is changed. 
We examined how many users of the diversity-increasing and decreasing-groups are reading in each cluster, extracted the differences in descending order, and confirmed the contents with human annotators.

Table \ref{table:cluster} shows the main contents of the extracted clusters.

The topics of clusters, which are often included in the diversity-increasing group, include international politics, economic information, and market information. 
On the other hand, clusters that have more in common in diversity-decreasing groups contain concentrated information on individual industries and companies and such specific topics as {\sl shogi}. This result suggests a positive correlation between interest in international politics and economic information and a widening range of interests.

\begin{table*}[ tbp]
\begin{center} 
\caption{Clusters that are mostly browsed by users belonging to each group
\newline
DI denotes diversity-increasing groups and DD denotes diversity-decreasing groups.
}\label{table:cluster}
\begin{tabular}{c}
\begin{minipage}[t]{1\hsize}
\begin{center} 
\hspace{1.6cm} [1] Mostly browsed by DI\\
\begin{tabular} {|c|c|c|}
\hline
{DI} $[\%]$&{DD} $[\%]$&{Main contents of cluster}\\
\hline
39.4&26.5&Politics\\
\hline
34.4&19.4&US market \& preparations for one's death\\
\hline
33.2&18.9&North Korea issues\\
\hline
31.5&18.3&Stock market information/updates\\
\hline
38.9&18.3&Business acquisition and withdrawal\\
\hline
\end{tabular} 
\end{center} 
\end{minipage}

\\
\begin{minipage}[c]{1\hsize}
\begin{center} 
\hspace{1.6cm} [2] Mostly browsed by DD\\
\begin{tabular} {|c|c|c|}
\hline
{DI} $[\%]$&{DD} $[\%]$&{Main cluster contents }\\
\hline
24.8&33.2&Market information about tech companies\\
\hline
15.5&26.2&Asian monetary policy\\
\hline
17.2&24.5&Yahoo! \\
\hline
14.1&22.0&The latest technology on automobiles\\
\hline
7.0&14.4&Famous shogi player\\
\hline
\end{tabular} 
\end{center} 
\end{minipage}

\end{tabular}
\end{center} 
\end{table*}

\subsection{Discussions}
\hspace{1em}
Although we identified few diversity-increasing groups, they contain relatively young users and more women, unlike the main users of Nikkei news who are men in their 40s and 60s. This observation suggests a positive correlation between age and a narrower range of interest, which is also consistent with a report in brain science \cite{berry2016aging} that suggests that cognitive flexibility declines with age.

In addition, there are clusters with different browsing ratios in the diversity increasing group and the diversity decreasing group. Users belonging to the diversity decreasing group read a lot articles that are of specific topics such as specific companies and industries and trendy topics such as {\sl shogi}.
On the other hand, for users belonging to the diversity increase group, the clusters contain a wide range of information such as international politics, economic information, and market information. 

Perhaps users in the diversity-decreasing group read interesting articles because they are very interested in a specific topic and much less interested in other topics. On the other hand, users in the diversity-increasing group are interested in such wide perspectives as financial markets and international politics, and their interests may be more easily targeted to such a broader range of topics.

\section{Modeling\label{sec:modeling}}

\subsection{Modeling of Interest\label{subsec:model}}
\hspace{1em}
Next, we propose an article selection model to clarify the influence of recommendation systems on behavioral changes.

In this research, we describe the behavior of general users in online media by a multi-agent model that expresses the behavior of all aspects of media.
Our model assumes that online media are news sites to which articles are added and updated every day. We observed the behavioral changes of users.

In our proposed model, for simplicity, we defined the unit time that corresponds to one day in the real world for updating the articles and user's interests and multiple articles selected by the users.
In this model, the displayed articles based on browsing history and user's interests are updated at most once a day.

The topics of the articles and the interests of users are represented by multiple categories.
Categories, which are assumed to have different properties, correspond to articles and users, and the topics of the user's interests and articles are determined by the distribution of categories.
Article $a_{i}$ continues to have an immutable property defined by the following vector constituted by category $c_{i}$:

\begin{eqnarray}
a_{i}=[c_{i1}, c_{i2}, ..., c_{in}].
 \label{eq:article}
\end{eqnarray}

User $u_{j}$ evaluates article $a_{i}$ based on the interest defined by the following vector and decides whether to browse it:
\begin{eqnarray}
u_{j}=[c_{j1}, c_{j2}, ..., c_{jn}].
 \label{eq:agent}
\end{eqnarray}

The evaluation of article $a_{i}$ of user $u_{j}$ is calculated by the following formula:
\begin{eqnarray}
P(u_{j}, a_{i})=u_{j}\cdot a_{i}.
 \label{eq:preference}
\end{eqnarray}

Depending on the articles browsed by the user, the interests of user $u$ are updated:
\begin{eqnarray}
u_{new}=w \cdot u_{old} + \sum_{i} k_{i} \sum_{j} c_{i,j}.
 \label{eq:agent_update}
\end{eqnarray}
In the formula, $w$ is a weight expressing how much of previous day's interest $u_{old}$ was stored, 
and $c_{i, j}$ is the value of category $c_{i}$ of article $a_{j}$. 
$k_{i}$ is a weight given to each category $c_{i}$ based on previous day's interest $u_{old}$. The weight expresses that the influence obtained from browsing is different based on the previous interests of each category.
For example, categories with low interest are hardly affected by browsing, and interesting areas are more likely to be strengthened by browsing interesting categories. 
In the experiment, we divided by the average value of categories, $k_{i} = 1 + r$, for the average and above, and $k_{i} = 1 - r$ for below average.

\subsection{Presentation and selection of articles}
\hspace{1em}
The top articles are presented to all users and individual articles are presented individually to each user. A fixed number of top articles are randomly selected. 
Individual articles are selected by each recommendation system.
The users select multiple articles a day from a set of top articles and individual articles and update interests.

\subsubsection{Selection of articles}
The user selects the articles to be browsed by elite and roulette selections based on the article's evaluation value calculated by formula \ref{eq:preference}. If the user selects an article whose evaluation value is less than threshold $th$, the user selects again, and if the user cannot find an article that exceeds threshold $th$ over a certain number of times, the user ends that turn's selection.

\begin{description}
 \item[Elite selection]: Choose a certain number from the top of the evaluation value among the presented articles.
 \item[Roulette selection]: Choose a certain number of articles by roulette selection from articles that were not chosen by elite selection.
\end{description}
Probability $\Pi_{i}$ that $a_{i}$ is selected is defined by the following formula:
\begin{eqnarray}
\Pi_{i}=\frac{P_{j}}{\sum_{j}P_{j}}.
 \label{eq:prob_kiji}
\end{eqnarray}
In this formula, $P_{j}$ is the evaluation value of article $a_{j}$ calculated by formula \ref{eq:preference}.

\subsubsection{Presentation of individual articles}
Individual articles are a set of articles selected by a recommendation system and randomly selected articles.
We compared two types of recommendation systems: a content-based recommendation based on the similarity of past browsing history and users, and a collaborative filtering recommendation that suggests items viewed by other similar users.

\subsubsection{Content-based recommendation}
In content-based recommendations, profile $Pf_{i}$ is calculated based on the browsing history of user $u_{i}$ using the following formula, and articles are selected in descending order of similarity between them and the profile:
\begin{eqnarray}
Pf_{i}=\sum_{j} c_{j}.
 \label{eq:profile}
\end{eqnarray}
Similar to \ref{eq:preference}, the similarity of profile $Pf_{i}$ and article $a_{j}$ is calculated:
\begin{eqnarray}
S=(Pf_{i}, a_{i})=u_{j}\cdot a_{i}.
 \label{eq:preference2}
\end{eqnarray}

At this time, instead of using interest vector $u_{i}$ of the users, we use profile $Pf_{i}$. Because identifying interest vector $u_{i}$ of the users from the recommendation system is relatively difficult, we have to predict interest vector $u_{i}$ of the users from the viewing history.

\subsubsection{Collaborative filtering}
With collaborative filtering, we recommend articles from users that have high similarity with other users as well as articles that have not been read yet. The similarity between the users is calculated based on the overlapping rate of the browsed articles. From their browsing history, two users, $u_{i}$，$u_{j}$, and the overlapping rate of their browsed articles, $A_{i}$，$A_{j}$, is obtained by the Simpson coefficient:

\begin{eqnarray}
Sim(u_{i},u_{j})=\frac{|A_{i} \cdot A_{j}|}{min(|A_{i}|,|A_{j}|)}.
 \label{eq:sim_agent}
\end{eqnarray}

First, we calculated the overlapping rate of browsed articles of all users, and select the users that have high similarity. Then, the user chooses the articles that have not been read yet but have been mostly read by the similar users. 

\subsubsection{Presentation without recommendations}
Without a recommendation system, actual users browse the articles displayed at the top of a website and those summarized as their favorites. Without recommendations, the day's most recent article is displayed and such highlighted articles depend on the browsing times of the users during the day. Considering the difference between their favorite types and daily browsing times, 100 articles were chosen by elite selection from the newest 1500 articles updated on that day, and 50 articles were randomly selected from 100.
To verify the simulation, we consider a method that presents all 5000 presentation article candidates for that day.

\subsection{Simulation procedure}
\hspace{1em}
As an initial setting, the categories of the interests of the articles and users are generated as uniform random numbers.
At each turn of the simulation, steps 1-4 are updated sequentially. For each turn, we evaluated the degree of the concentration of the categories of the interests of the users and observed the changes in them.

\begin{enumerate}
 \item Update new articles.
 \item Update the presented articles:
 \begin{itemize}
 \item Update the top articles.
 \item Update the recommended articles.
  \begin{itemize}
  \item Calculate the user similarities and select similar users (collaborative filtering).
  \item Calculate users and article similarities (content-based recommendation).
 \item Select recommended articles.
\end{itemize}
 \item Update individual random articles. 
\end{itemize}
 \item Select articles by users.
 \item Update the user's interest.
\end{enumerate}

\section{Verification of Model}

\subsection{Simulation settings}
\hspace{1em}
Next we compared our proposed model and the Nikkei news data that were analyzed in Section \ref{sec:analysis_feature}.
In this verification experiment, we compared the changes in the cluster entropy of the Nikkei news and in the entropy of the simulation's interest categories. If these changes are similar, they are an appropriate article selection model of online media.

We selected articles depending on the proposed model and compared the change of diversity with Nikkei's electronic version. 
As an indicator of diversity, like in Section \ref{sec:analysis_behav}, we evaluated the diversity of the browsing behavior based on the degree of concentration of the categories of the articles that were read. The degree of category concentration was calculated by the following formula as information entropy:

\begin{eqnarray}
H(u)=-\sum_{i}p_{i} \cdot log p_{i}.
 \label{eq:entropy2}
\end{eqnarray}
In this formula, $p_{i}$ represents each user's existence probability for each category $c_{i}$.

A user with high category entropy is reading articles composed of various categories, and a user with low category entropy is intensively reading articles composed of a specific category.
We also compared whether the maximum category for each user changed before and after the period to analyze changes in users' interests.

In this simulation, we changed and compared the recommendation system.
Table \ref {table:simulation} shows the detailed settings of the simulation.
For a breakdown of the daily presentation articles, the top article is fixed at 50 articles, and the remaining 50 articles were suggested by each recommendation system.
We set the following four scenarios.
\begin{itemize}
  \item ContentBase : 50 articles selected by content-based recommendation.
  \item Collaborative : 50 articles selected by collaborative filtering recommendation
  \item NonRecommendation : 50 articles selected randomly from 100 articles by elite from the latest 1500 articles.
  \item All : All 5000 presented articles for that day, and the only scenario: All is different from the other scenarios respect to the number of articles presented per day.
\end{itemize}

This simulation period lasted for 45 turns (45 days) from June 1, 2017 to July 15, 2017. 
We did this simulation 20 times under identical conditions. The results described below are averages.

\begin{table} [ tbp]
\begin{center} 
\caption{Detailed simulation settings}
\label{table:simulation}
\begin{tabular} {|c|c|c|}
\hline
Simulation period&45\\
\hline
Number of simulations&20\\
\hline
Number of users&1000\\
\hline
Number of categories&20\\
\hline
Number of presented candidate articles per day&5000\\
\hline
Number of articles updated per day&1500\\
\hline
Number of articles presented per day&100\\
\hline
Number of articles read per day&10\\
\hline
Elite selected articles per day&3\\
\hline
Number of high similarity users for collaborative filtering&20\\
\hline
Threshold value of evaluation value$th$&0.055\\
\hline
Weight of interest of previous day $w$&5\\
\hline
Category weight $r$&0.5\\
\hline
\end{tabular} 
\end{center} 
\end{table}

\subsection{Results}
\hspace{1em}
Table \ref{table:simulation_result} shows the simulation results when the recommendation system was changed.
Among the results of each scenario, the average category entropy had a significant difference at a significance level of 1\%. 
For the proportion of users with the largest category change, {\it All-NonRecommendation}, {\it NonRecommendation-ContentBase} had a significant difference at a significance level of 5\%, and {\it ContentBase-Collaborative}, {\it Collaborative-All}, {\it NonRecommendation-Collaborative} had a significant difference at a significance level of 1\%.
There was no significant difference between {\it All} and {\it ContentBase}.

\begin{table} [ tbp]
\begin{center} 
\caption{Simulation results
\newline
ACE denotes the average of the interest category entropy, MCC denotes the max category changes, FP denotes the first period, and LP denotes the last period.
}
\label{table:simulation_result}
\begin{tabular} {|c|c|c|c|}
\hline
\multicolumn{1}{|c|}{Scenario} &\multicolumn{2}{c|}{ACE}
      & \multicolumn{1}{c|}{MCC}\\ 
\multicolumn{1}{|c|}{} &\multicolumn{1}{c}{FP} &\multicolumn{1}{c}{LP}
      & \multicolumn{1}{|c|}{}\\ \hline 
ContentBase&3.63&1.21&11.1\\
Collaborative&3.63&2.80&12.6\\
All&3.63&2.54&11.2\\
NonRecommendation&3.63&1.50&10.1\\
\hline
\end{tabular} 
\end{center} 
\end{table}

\subsubsection{Change in entropy}
\begin{figure} [ tbp]
\begin{center} 
\includegraphics[width=0.8\linewidth]{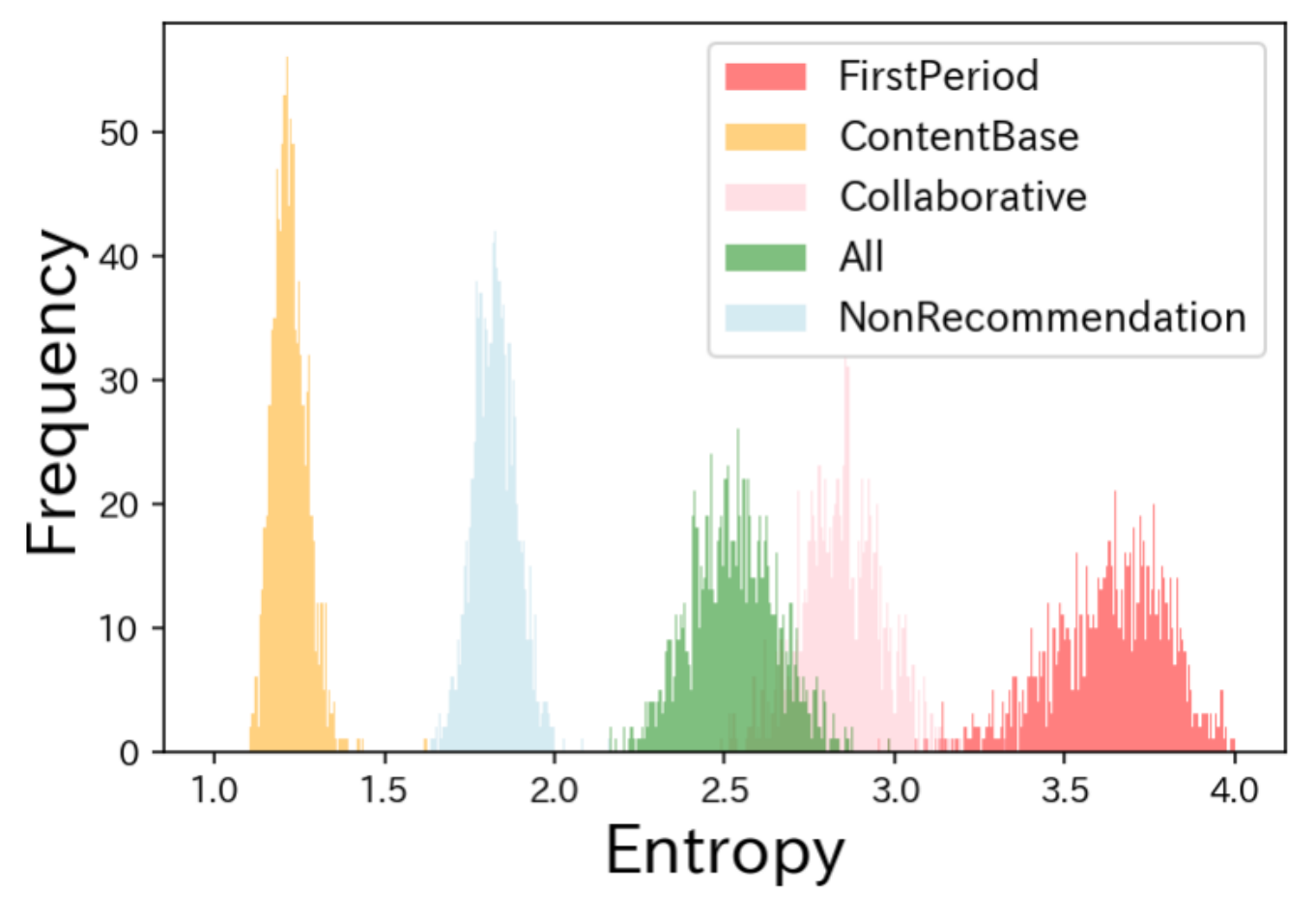} 
\caption{Changes in interest category entropy}
\label{simulation_entropy}
\end{center} 
\end{figure}

Figure \ref {simulation_entropy} shows the change in the interest category entropy at the start and end of the simulation in each setting.
The interested category entropy is identical to that shown in Fig. \ref{entropy} during the period. As the distribution moves to the left, entropy tends to decrease. We confirmed that since the simulation is done through the period, the diversity of the browsed articles tended to decrease. From the above result, the proposed model is valid because it can present the same article selection as Nikkei.

We are also aware that compared to a recommendation system, recommendation based on contents ({\it ContentBase}) reduces the diversity the most. 
We notice that the diversity of recommendations based on collaborative filtering ({\it Collaborative}) has the smallest decrease in diversity, and the presentation method ({\it NonRecommendation} and {\it All}), which assumes a case without recommendations, is located in the middle.

Between {\it NonRecommendation} and {\it All}, since only {\it NonRecommendation} gives a favorite genre to a candidate, diversity decreases more.
When all the articles are presented ({\it All}), diversity decreases.
When the user updates its interested category entropy, the interest presented in the past browsed articles tends to weigh more than the actual achieving individualization in Nikkei, which reduces the diversity. Hence our proposed model shows its advantage at this time.

Since Nikkei currently has no recommendation system, it corresponds to the middle of {\it NonRecommendation} and {\it All}, which is a presentation method that assumes a case without recommendations. 
This is because the presented articles correspond to determining whether to browse based on article titles in the real world. Although interesting articles spill over genres, articles that generally belong to a single genre (uninteresting) are not recognized, and not even their titles are confirmed. In other words, some choices are made at the stage of the recognition of articles, even in the present situation where individualization is not substantially done. Therefore, in reality, the reduction of diversity is likely to occur without an individualized recommendation system.

Next, concerning the influence of such recommendation systems as content-based recommendations and collaborative filtering, the former only presents articles that were valued from browsing history, and since positive feedback is strongly applied, the diversity fell the most. Recommending articles with high similarity based on contents is not preferable for suppressing the decrease of diversity. In addition, collaborative filtering, which decreases diversity less than {\it NonRecommendation}, is appropriate for suppressing filter bubbles.
Although diversity decreases in any case after a certain period of time, collaborative filtering decreases diversity less than {\it NonRecommendation}, and this result is also consistent with the MovieLens analysis by Nguyen et al. \cite{nguyen2014exploring}.

From this, a method that recommends articles with high similarity based on contents is not desired for avoiding a decrease of diversity. Collaborative filtering, which suppresses a decrease in diversity more than without recommendations, is appropriate for suppressing filter bubbles. 

\subsubsection{Change in interest category}
\begin{figure}[ tbp]
\begin{tabular}{cc}

\begin{minipage}[c]{0.5\hsize}
\includegraphics[width=1\linewidth]{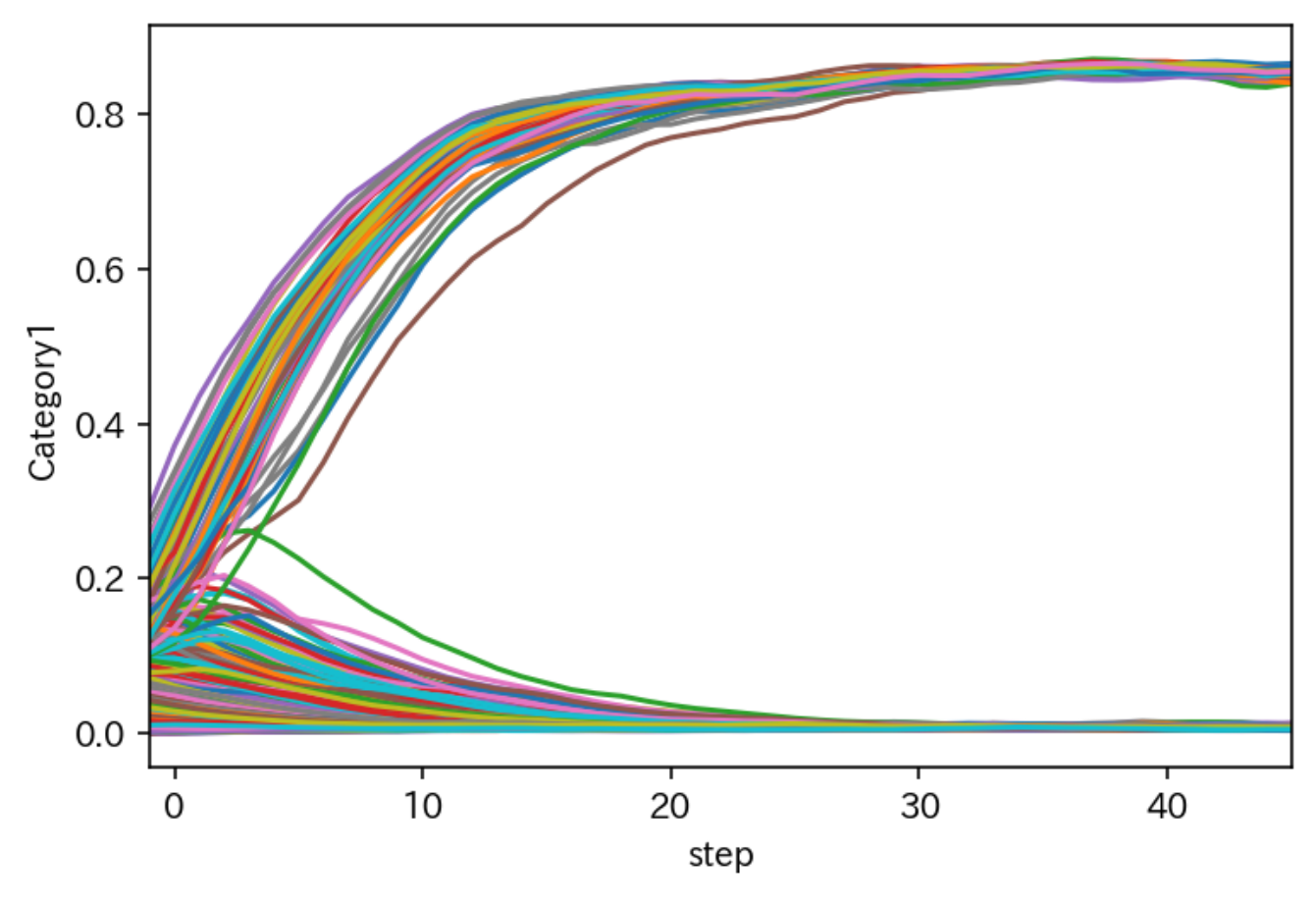} 
\hspace{1.6cm} [1] ContentBase
\end{minipage}
\begin{minipage}[c]{0.5\hsize}
\includegraphics[width=1\linewidth]{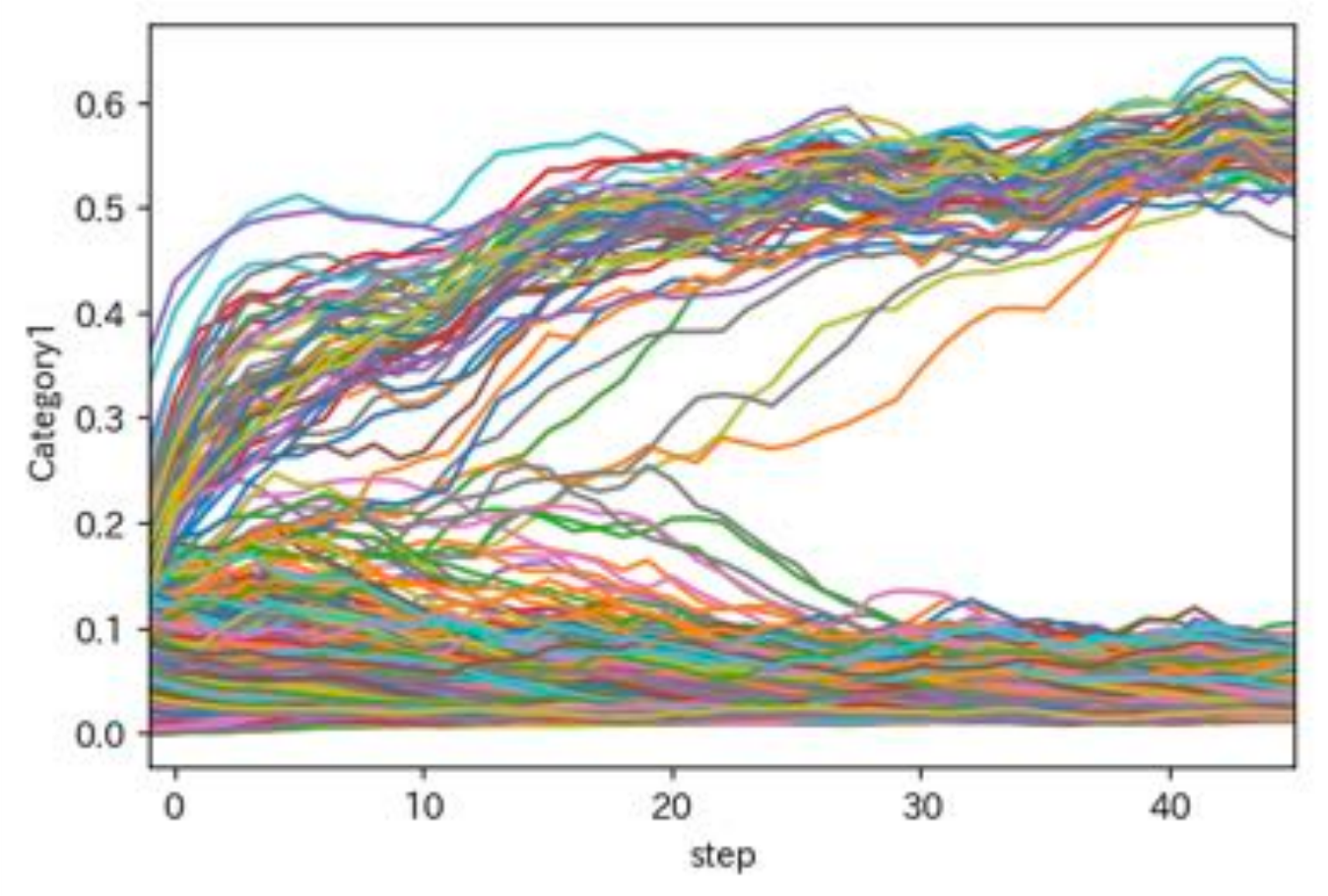} 
\hspace{1.6cm} [2] Collaborative
\end{minipage}

\\
\begin{minipage}[t]{0.5\hsize}
\includegraphics[width=1\linewidth]{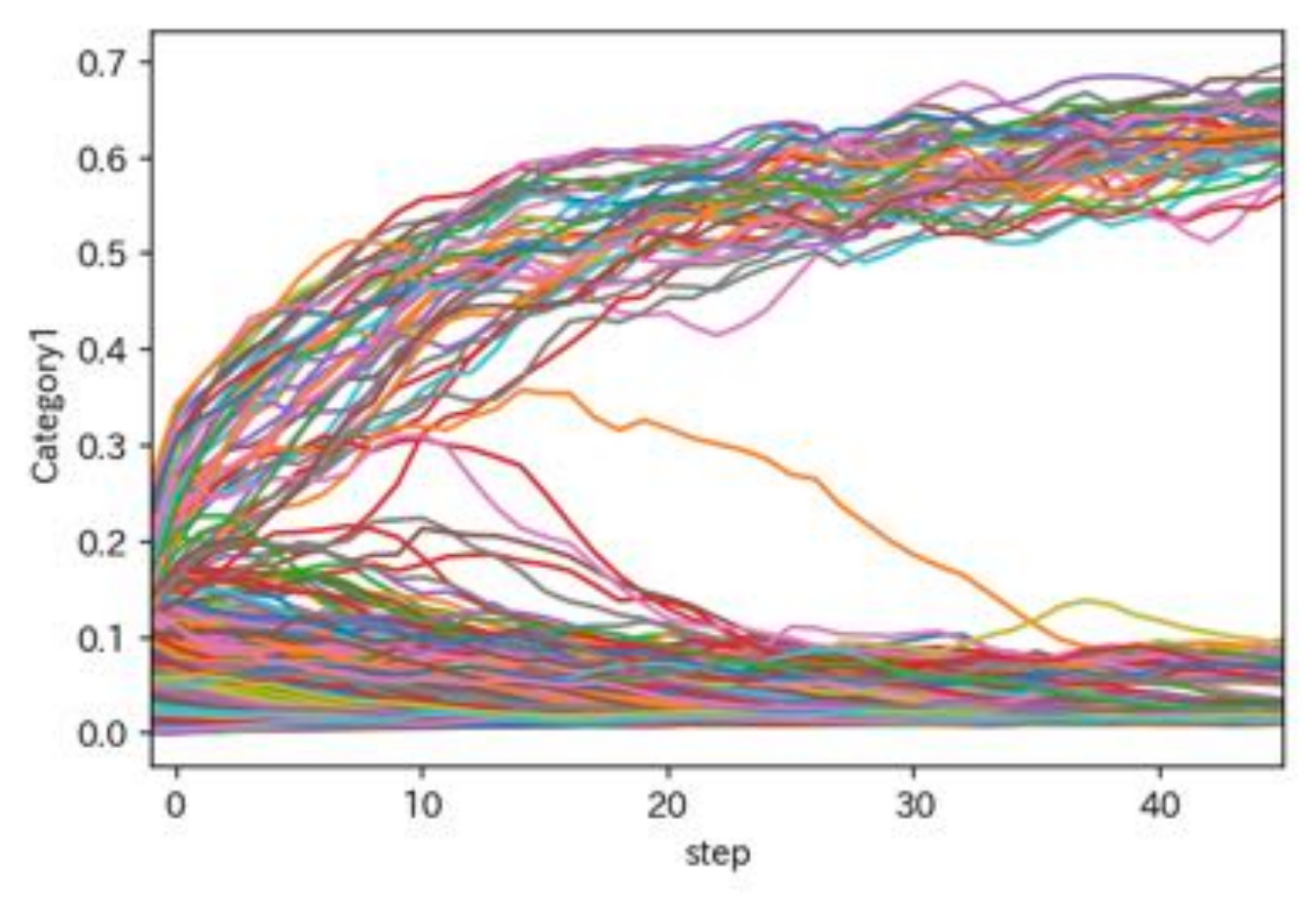} 
\hspace{1.6cm} [3] All
\end{minipage}
\begin{minipage}[t]{0.5\hsize}
\includegraphics[width=1\linewidth]{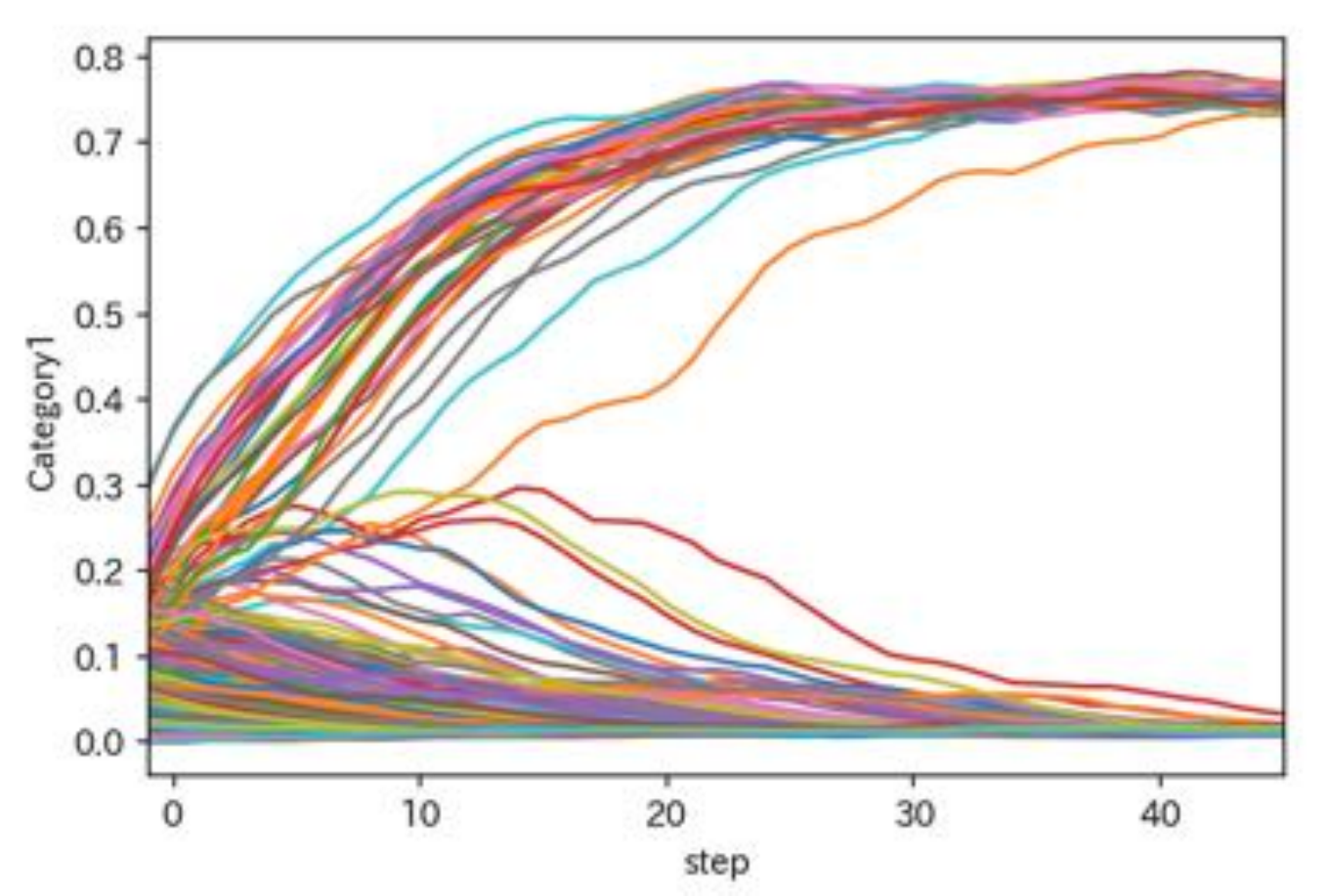} 
\hspace{1.6cm} [4] NonRecommendation
\end{minipage}

\end{tabular}
\caption{Changes of interest category 1 for each user in each scenario}\label{log}
\end{figure}

We identified a slightly different tendency from the changes in interest category entropy. However, the collaborative filtering rate changed the most and seems to indicate not only a smaller diversity decrease but also that user interest is likely to change. Fig. \ref{log} shows the changes of interest category 1 for each user in each scenario. The horizontal axis represents the number of simulation steps, and the vertical axis represents the proportion of interest category 1 in the interest category of each user.
At the beginning, the category bias is small and distributed between $0-0.3$. 
At the end of the period, although bias is occurring in every case, the bias in the case of recommendations based on content ({\it ContentBase}) is remarkable and the convergence is also fast. {\it NonRecommendation} also has a large bias. Even at the end of the simulation by collaborative filtering, the proportion of category 1 of the group with a larger category 1 is $0.5-0.6$, which indicates that the deviation is weaker than the others.

From these results, the maximum category is easily enhanced, and at the simulation's end many users account for more than half of the interest category. Such a maximum category is likely to be strengthened, reducing diversity.
In other words, a recommendation method that can suppress the strengthening of the maximum category is effective for developing a recommendation system that can suppress a decrease in diversity.

\section{Conclusions}
\hspace{1em}
We analyzed behavioral changes prior to recommendations by network clustering and found that user attributes and cluster contents are different among users with different behavioral changes. The proportion of users younger than 40 and women was relatively large in the diversity-increasing group.

We also proposed an article selection model and clarified the influence of recommendation systems on behavioral changes. Simulation results showed that diversity generally decreases, but collaborative filtering can suppress its decrease more than without recommendations. In addition, we found that the maximum category is easily strengthened, which is one factor that causes less diversity.

Future research will continue to analyze the factors that lead to behavioral changes for developing effective recommendation systems. 
In addition, we will develop recommendation systems based on these results and improve our models.
In the development and experiments of recommendation systems, we will use our findings and compare recommendation systems and behavioral changes.

\bibliographystyle{abbrv}
\bibliography{wi}

\begin{thebibliography}{10}

\bibitem{ahmed2014agent}
L.~Ahmed and A.~Abhari.
\newblock Agent-based simulation of twitter for building effective recommender
  system.
\newblock In {\em Proceedings of the 17th Communications \& Networking
  Simulation Symposium}, page~5. Society for Computer Simulation International,
  2014.

\bibitem{baba2015classification}
S.~Baba, F.~Toriumi, T.~Sakaki, K.~Shinoda, S.~Kurihara, K.~Kazama, and
  I.~Noda.
\newblock Classification method for shared information on twitter without text
  data.
\newblock In {\em Proceedings of the 24th International Conference on World
  Wide Web}, pages 1173--1178. ACM, 2015.

\bibitem{bakshy2015exposure}
E.~Bakshy, S.~Messing, and L.~A. Adamic.
\newblock Exposure to ideologically diverse news and opinion on facebook.
\newblock {\em Science}, 348(6239):1130--1132, 2015.

\bibitem{berry2016aging}
A.~S. Berry, V.~D. Shah, S.~L. Baker, J.~W. Vogel, J.~P. O'Neil, M.~Janabi,
  H.~D. Schwimmer, S.~M. Marks, and W.~J. Jagust.
\newblock Aging affects dopaminergic neural mechanisms of cognitive
  flexibility.
\newblock {\em Journal of Neuroscience}, 36(50):12559--12569, 2016.

\bibitem{blondel2008fast}
V.~D. Blondel, J.-L. Guillaume, R.~Lambiotte, and E.~Lefebvre.
\newblock Fast unfolding of communities in large networks.
\newblock {\em Journal of statistical mechanics: theory and experiment},
  2008(10):P10008, 2008.

\bibitem{cosley2003seeing}
D.~Cosley, S.~K. Lam, I.~Albert, J.~A. Konstan, and J.~Riedl.
\newblock Is seeing believing?: how recommender system interfaces affect users'
  opinions.
\newblock In {\em Proceedings of the SIGCHI conference on Human factors in
  computing systems}, pages 585--592. ACM, 2003.

\bibitem{fleder2009blockbuster}
D.~Fleder and K.~Hosanagar.
\newblock Blockbuster culture's next rise or fall: The impact of recommender
  systems on sales diversity.
\newblock {\em Management science}, 55(5):697--712, 2009.

\bibitem{kiousis2001public}
S.~Kiousis.
\newblock Public trust or mistrust? perceptions of media credibility in the
  information age.
\newblock {\em Mass Communication \& Society}, 4(4):381--403, 2001.

\bibitem{lee2002intelligent}
W.-P. Lee, C.-H. Liu, and C.-C. Lu.
\newblock Intelligent agent-based systems for personalized recommendations in
  internet commerce.
\newblock {\em Expert Systems with Applications}, 22(4):275--284, 2002.

\bibitem{pariserWrong}
G.~Linden.
\newblock Eli pariser is wrong.
\newblock available at:
  \url{http://glinden.blogspot.com/2011/05/eli-pariser-is-wrong.html}, 2011.
\newblock Accessed: May 3 2018.

\bibitem{linden2003amazon}
G.~Linden, B.~Smith, and J.~York.
\newblock Amazon. com recommendations: Item-to-item collaborative filtering.
\newblock {\em IEEE Internet computing}, 7(1):76--80, 2003.

\bibitem{matsuo2005social}
Y.~Matsuo, H.~Tomobe, H.~Nakashima, M.~Ishizuka, et~al.
\newblock Social network extraction from the web information.
\newblock pages 46--56, 2005.

\bibitem{mooney2000content}
R.~J. Mooney and L.~Roy.
\newblock Content-based book recommending using learning for text
  categorization.
\newblock In {\em Proceedings of the fifth ACM conference on Digital
  libraries}, pages 195--204. ACM, 2000.

\bibitem{nguyen2014exploring}
T.~T. Nguyen, P.-M. Hui, F.~M. Harper, L.~Terveen, and J.~A. Konstan.
\newblock Exploring the filter bubble: the effect of using recommender systems
  on content diversity.
\newblock In {\em Proceedings of the 23rd international conference on World
  wide web}, pages 677--686. ACM, 2014.

\bibitem{pariser2011filter}
E.~Pariser.
\newblock {\em The filter bubble: What the Internet is hiding from you}.
\newblock Penguin UK, 2011.

\bibitem{sarwar2001item}
B.~Sarwar, G.~Karypis, J.~Konstan, and J.~Riedl.
\newblock Item-based collaborative filtering recommendation algorithms.
\newblock In {\em Proceedings of the 10th international conference on World
  Wide Web}, pages 285--295. ACM, 2001.

\bibitem{schafer1999recommender}
J.~B. Schafer, J.~Konstan, and J.~Riedl.
\newblock Recommender systems in e-commerce.
\newblock In {\em Proceedings of the 1st ACM conference on Electronic
  commerce}, pages 158--166. ACM, 1999.

\bibitem{schein2002methods}
A.~I. Schein, A.~Popescul, L.~H. Ungar, and D.~M. Pennock.
\newblock Methods and metrics for cold-start recommendations.
\newblock In {\em Proceedings of the 25th annual international ACM SIGIR
  conference on Research and development in information retrieval}, pages
  253--260. ACM, 2002.

\bibitem{senecal2004influence}
S.~Senecal and J.~Nantel.
\newblock The influence of online product recommendations on consumers’ online
  choices.
\newblock {\em Journal of retailing}, 80(2):159--169, 2004.

\bibitem{uchida2017evaluation}
K.~Uchida, F.~Toriumi, and T.~Sakaki.
\newblock Evaluation of retweet clustering method classification method using
  retweets on twitter without text data.
\newblock In {\em Proceedings of the International Conference on Web
  Intelligence}, pages 187--194. ACM, 2017.

\end{thebibliography}
\end{document}